\documentclass[reprint,amsmath,amssymb,aps,]{revtex4-2}

\usepackage{graphicx}
\usepackage{dcolumn}
\usepackage{bm}

\usepackage{blindtext}
\usepackage{xcolor}
\usepackage{cancel}
\usepackage[normalem]{ulem}

\begin{document}

\preprint{APS/123-QED}

\title{Surface instabilities generated by a slider pulled across a granular bed}
\author{Antoine Dop}
\email{antoine.dop@ens-lyon.fr}
\affiliation{ENSL, CNRS, Laboratoire de Physique, F-69342 Lyon, France.\\
}
\author{Valérie Vidal}
\email{valerie.vidal@ens-lyon.fr}
\affiliation{ENSL, CNRS, Laboratoire de Physique, F-69342 Lyon, France.\\
}

\author{Nicolas Taberlet}
\email{nicolas.taberlet@ens-lyon.fr}
\affiliation{Univ Lyon, UCBL, ENSL, CNRS, Laboratoire de Physique, F-69342 Lyon, France.
}
\altaffiliation[corresponding author: ]{nicolas.taberlet@ens-lyon.fr}

%\date{\today}
\author{(Received 7 March 2023; accepted 1 July 2023; published 9 August 2023)}

\begin{abstract}

We report an instability of a slider slowly dragged at the surface of a granular bed in a quasistatic regime. The boat-shaped slider sits on the granular medium under its own weight and is free to translate vertically and to rotate around the pitch axis while a constant horizontal speed is imposed. For a wide range of parameters (mass, length, shape, velocity) a regular pattern of peaks and troughs spontaneously emerges as the slider travels forward. This instability is studied through experiments using a conveyor belt and by means of two-dimensional discrete elements method simulations. We show that the wavelength and amplitude of the pattern scale as the length of the slider. We also observe that the ripples disappear for low and high masses, indicating an optimal confining pressure. The effect of the shape, more specifically the inclination of the front spatula, is studied and found to drastically influence both the wavelength and the amplitude. Finally, we show that the mechanical details (friction, cohesion) of the contact point between the slider and the pulling device is critical and remains to be fully understood.

\end{abstract}

\maketitle

\section{\label{sec:introduction}Introduction}

Granular materials exhibit a wide variety of complex behaviors including the formation of patterns for which the grains can spontaneously arrange with a typical length scale much larger than the size of individual grains~\cite{Aranson2006, Ristow2000}.
Such pattern formation can be triggered by interaction with a fluid flow, such as aeolian dunes and ripples \cite{Charru2013}, underwater ripples \cite{Charru2013,Kennedy1969}, or shear banding~\cite{Varga2019}. Another example is the fingering instability which can emerge from surface tension \cite{Sandnes2011}, drying~\cite{Yamazaki2000} or viscosity~\cite{Homsy1987}. Other pattern formation can also happen with grains vibrated either vertically or horizontally~\cite{Umbanhowar1996}, placed in a rotating drum \cite{Seiden2011}, in a simple flow down an inclined plane~\cite{Pouliquen1997a,Forterre2003,Pouliquen1997b,Goldfarb2002}, or due to injecting grains through a matrix of larger grains~\cite{Pinto2007}.

Starting from a flat granular bed, a corrugated topography with a fixed wavelength can be obtained when repeatedly dragging an inclined plate (whose angle is fixed) or rolling a wheel on top of the grains~\cite{Bitbol2009,Percier2013}. This effect known as washboard road instability can also be observed after a single passage \cite{Hewitt2012a}, and exists in other materials such as viscoplastic fluids \cite{Hewitt2012a,Hewitt2012b}.
Importantly, this instability develops only if the velocity is beyond a threshold \cite{Both2001}, and any initial perturbation in the bed is erased below this critical velocity. 

Pulling a mass attached to a spring on top of a substrate is the historical experiment of dry friction, and is still used today \cite{Vidal2019}. Such a setup has previously been used on a granular bed~\cite{Nasuno1998}, leading to three force regimes depending on the values of the mass, the spring stiffness and the pulling speed. Additionally, some experiments included transverse mechanical vibrations which had the effect to kill the stick-slip regime at a certain vibration speed threshold \cite{Vidal2019}. Vibrations are analogous to temperature in granular materials and are known to play a crucial role in granular flow and properties \cite{Capozza2009}.

\begin{figure*}[t]
\includegraphics[width = 1.5\columnwidth]{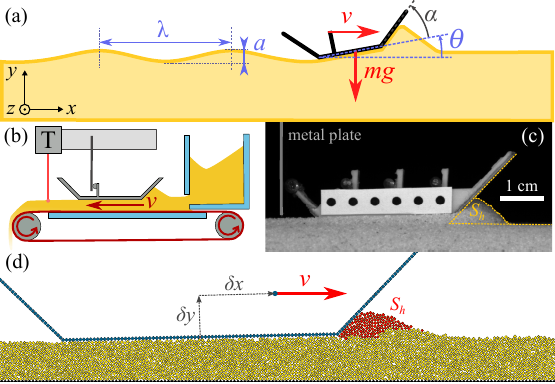}
\caption{\label{fig:sketch}
(a) Sketch of the instabilities generated by a slider over a granular bed (the real wavelength is much larger than the slider length). The slider (mass $m$) is pulled at a constant speed $v$ via a rigid metallic plate. Its pitch angle is denoted $\theta$. 
(b) Sketch of the experimental setup in the threadmill configuration, with the laser telemeter T measuring the height $h$ in the slider wake. The belt rests on a rigid horizontal plate, ensuring a flat horizontal bottom.
(c) Image of the side view of the experiment displaying the heap size $S_h$ [$m=18$~g, $l=30$~mm, $\alpha=45$°]. 
(d) Snapshot of a two-dimensional discrete elements method (2D DEM) simulation [$m=0.15$~g, $l=48$~mm, $\alpha=45$°]. The figure only displays a small portion of the total simulation box, which length is of about 3~m. The grains moving faster than $62.5\%$ of $v$ are colored in red and correspond to the heap surface $S_h$ (see text).}
\end{figure*}

In this paper we investigate an instability involving slowly pulling a slider across a granular bed. It is reminiscent of the washboard road instability, with several distinct features. First, the slider has two degrees of freedom (rotation around the pitch axis and vertical translation, see Fig.~\ref{fig:sketch}), and we monitor the state of the granular bed after a unique passage of the slider (as opposed to the repeated passages in the case of the washboard road instability). 
Doing so may result in the formation of ripples with a well defined wavelength. 
Second, this instability occurs in the quasistatic regime and shows no dependence on the pulling speed. 
These differences suggest that the instability that we present is original and, to our knowledge, has never been reported.

The paper is organized as follows. Our experimental setup is described in Sec.~\ref{sec:exp_methods}, and our simulations methods in Sec.~\ref{sec:num_methods}. We then present experimental and numerical results in Sec.~\ref{sec:results}, mainly focusing on the role of the mass and geometry of the slider on the amplitude and wavelength of the pattern. We then discuss these results and mention the forces involved in the problem.

\section{\label{sec:methods}Methods}

Figure~\ref{fig:sketch}(a) displays the surface instability (amplitude, $a$, and wavelength, $\lambda$) generated by a slider of mass, $m$, pulled at velocity, $v$ across a granular bed. Here we choose to pull the slider via a very rigid metal plate, ensuring a displacement at constant speed (see discussion in Sec.~\ref{sec:discussion}). $g$ is the gravitational acceleration, and $y$ is the vertical position of the middle of the bottom plate on the slider, $\theta$ is the slider angle with respect to the horizontal, and $h$ is the thickness of the granular layer left after the passage of the slider. 
In the following, we consider the difference $y-y_0$, where $y_0 = y(t=0)$ is the altitude of the slider at the beginning of the experiment, when it is resting on top of the granular bed. We also refer to topography as $(h-h_0)$, the height difference respect to the initial granular bed height, $h_0$. Finally, note that the slider angle is relative to the horizontal axis, with the slider pointing upwards when $\theta>0$.
No stick-slip motion is observed as the metal plate pulling the slider has a high stiffness. Therefore, all quantities can be easily plotted either as a function of time, or instead as a function of the traveled distance, $x$.
The next subsections present the experimental and numerical methods used. 

\begin{table*}[t]
\caption{\label{tab:table1}Range of parameters used in experiments and simulations. Since simulations are 2D, the 2D equivalent of experimental parameters is given when relevant. To do so, a scaling factor $d_p/w$ is applied to quantities which scale as the dimension along $z$, the direction transverse to the slider horizontal and vertical motion. The confining pressure is defined as $P=mg/(lw)$.}

\begin{ruledtabular}
\begin{tabular}{ccccc}
 Parameter&Description&Experimental&2D equivalent ($\times d_p/w$)&Numerical\\ \hline
 $d_p$&Grain diameter&$430 \pm 100$~$\mathrm{\mu}$m&&$500 \pm 100$~$\mathrm{\mu}$m\\
 $\rho_p$&Grain density&$(2.63 \pm 0.1) \times 10^3$~kg.m$^{-3}$&&$1.27\times 10^3$~kg.m$^{-3}$ \footnote{Considering that 2D grains are disks of width equal to diameter $d_p$, the mass of one grain divided by its surface is $0.637$ kg.m$^{-2}$}\\
  $h_0$&Initial thickness of the granular bed&10~mm&&$8.4$~mm\\
  $v$&Pulling speed&[1 - 100] mm/s&&[1 - 1000] mm/s \footnote{In all simulations shown in the following, $v=40$~mm/s}\\
 $m$&Mass of slider&[$10$ - $100$] g&[$0.1$ - $1$] g&[$1.25$ - $25$] $\times 10^{-2}$ g\\
 $l$&Slider length&[30 - 100] mm&&[12 - 180] mm\\
 $w$&Slider width&[40 - 100] mm&$d_p$&$d_p$\\
 $P$&Confining pressure&[40 - 830] Pa&&[3 - 330]~Pa\\
 $\alpha$&Slider spatula front angle&[30 - 60]°&&[5 - 120]°\\
  $\delta_x$&Horizontal force position\footnote{Relative to the center of the bottom plate of the slider}&[-30 +30] mm&&[-50 +50] mm\\
  $\delta_y$&Vertical force position\footnotemark[3]&[5 - 10] mm&&[-20 +20] mm\\
\end{tabular}
\end{ruledtabular}
\end{table*}

\subsection{\label{sec:exp_methods}Experimental setup}

The experimental setup is represented in Fig.~\ref{fig:sketch}(b).
Instead of dragging a slider on a static granular bed (whose length can only be limited), we decided to perform experiments on a conveyor belt which carries a uniform layer of sand at a constant speed, $v$. 
In all our experiments (and numerical simulations), the velocity $v$ is chosen to be small enough ($v/\sqrt{g d_p } \ll 1$) to ensure that the regime is quasistatic. In other words, the time it takes for a grain to settle down under its own weight is much shorter than the typical time it takes for a grain to travel horizontally by a distance $d_p$ when pushed forward by the slider. 
In this range of velocity (see Table~\ref{tab:table1}), the characteristics of the instability are independent of $v$. 
During any experiment, if the conveyor belt is stopped and started again, then the evolution of all quantities remain unchanged, confirming the quasistatic regime. Hence, the traveled distance, $x$, appears to be a more relevant variable than time to plot all physical quantities. 

The granular material is washed natural sand made of polydisperse grains of density $\rho = 2630 \pm 100$ kg.m$^{-3}$ and mean diameter $d_p = 430 \pm 100~\mathrm{\mu m}$ (median 420~$\mathrm{\mu m}$). A rectangular slit oriented along $z$, the direction transverse to the view in Fig.~\ref{fig:sketch}(b), at the bottom of the feeding tank,  allows the grains to flow on a $150$~mm-wide conveyor belt. The belt is held under a slight tension by two cylinders [gray disks in Fig.~\ref{fig:sketch}(b)], one of which is driven by a DC motor and gearbox (Crouzet 80 807 0Y00250Z) to provide a constant speed, $v$. As it is not perfectly rigid, the belt is resting on a $1$~cm-thick horizontal acrylic plate thus ensuring a flat, horizontal bottom.

The sliders are three-dimensionally (3D) printed and designed on SolidWorks, allowing precise control over the geometry. They are manufactured in polylactic acid (PLA) using Fused Deposition Modeling (FDM) 3D printing. They consist in a rectangular horizontal bottom plate and two inclined plates on the front and rear [Fig.~\ref{fig:sketch}(c)].
Sliders of various length, $l$, width, $w$, and front spatula angle, $\alpha$, were printed (Sec.~\ref{sec:angle}). Walls on either side prevent sand from getting on the slider and changing its mass over time. A monolayer of grains from the same batch as the granular bed is glued to the slider bottom surface using double-sided tape to ensure a grain-grain contact. The slider can be weighed using evenly distributed steel beads placed on its bottom plate so that its center of gravity remains vertical to its geometrical center. By doing so, its mass, $m$, can be varied between 10 and 100~g (Table~\ref{tab:table1}).
At the beginning of an experiment, the slider is carefully placed over the grains surface before the conveyor belt is started. A fixed rigid metal plate comes in contact with a horizontal cylinder glued to a vertical pillar (whose location can vary, see Fig.~\ref{fig:sketch}(c)) and pushes the slider forward. Using a cylinder as the contact surface ensures that the motion remains planar (in the plane shown in Fig.~\ref{fig:sketch}), i.e., with no roll or yaw of the slider. 

A side view of the experiment is recorded using a camera (Basler acA2040-90um with a Japan Zoom Lens 18-108/2.5).
Six black dots are printed on the side of the slider 
as seen in Fig.~\ref{fig:sketch}(c) in order to track its vertical position, $y$, and inclination, $\theta$. Moreover, the amount of grains plowed in front of the slider is measured through image processing (see Appendix~\ref{sec:appendix}).  A telemeter (Micro-epsilon, ILD1302-20) is placed behind the slider and measures the height, $h(x)$, of the granular bed left in its wake (Fig.~\ref{fig:sketch}(b), left). Additionally, a temperature and humidity sensor (TE Connectivity, HTM2500LF) is located inside the tank, immersed in the grains. All experiments were performed in ambient conditions.

\subsection{Numerical simulations}
\label{sec:num_methods}

The discrete elements method (DEM) has been used extensively to simulate the behavior of deformable (yet stiff) disks~\cite{Cundall1979, Percier2011, Sautel2021, Guo2015}.
This method relies on the computation of contact forces depending on the relative position and velocity of grains, and the numerical integration of the equations of motion. The normal component of the contact force is the sum of a repulsive term proportional to the overlap
with a numerical stiffness $k_N = 10^3$, and a viscous damping term.  
The numerical value of $k_N$ is several orders of magnitude lower than it should be to be match the Young modulus of sand particles. This choice allows one to take a significantly higher value of the timestep since collision times are directly related to $k_N$ (as well as viscous dissipation and the mass of grains). This numerical trick has been widely used in similar simulations and improves greatly the computing time without affecting the flow properties of the grains \cite{GDR2004,Lommen2014}. This low value of $k_N$ alters the propagation of sound waves in the granular packing, but this property is not relevant in the present study. Note that one should expect the value of the viscous damping coefficient to have no important effect in a dense quasistatic regime. 
The tangential component is given by the Cundall model~\cite{Cundall1979}, which models solid friction with a memory effect. The corresponding solid-friction coefficient is set to $0.8$. 

\begin{figure*}[t]
\includegraphics[width=\textwidth]{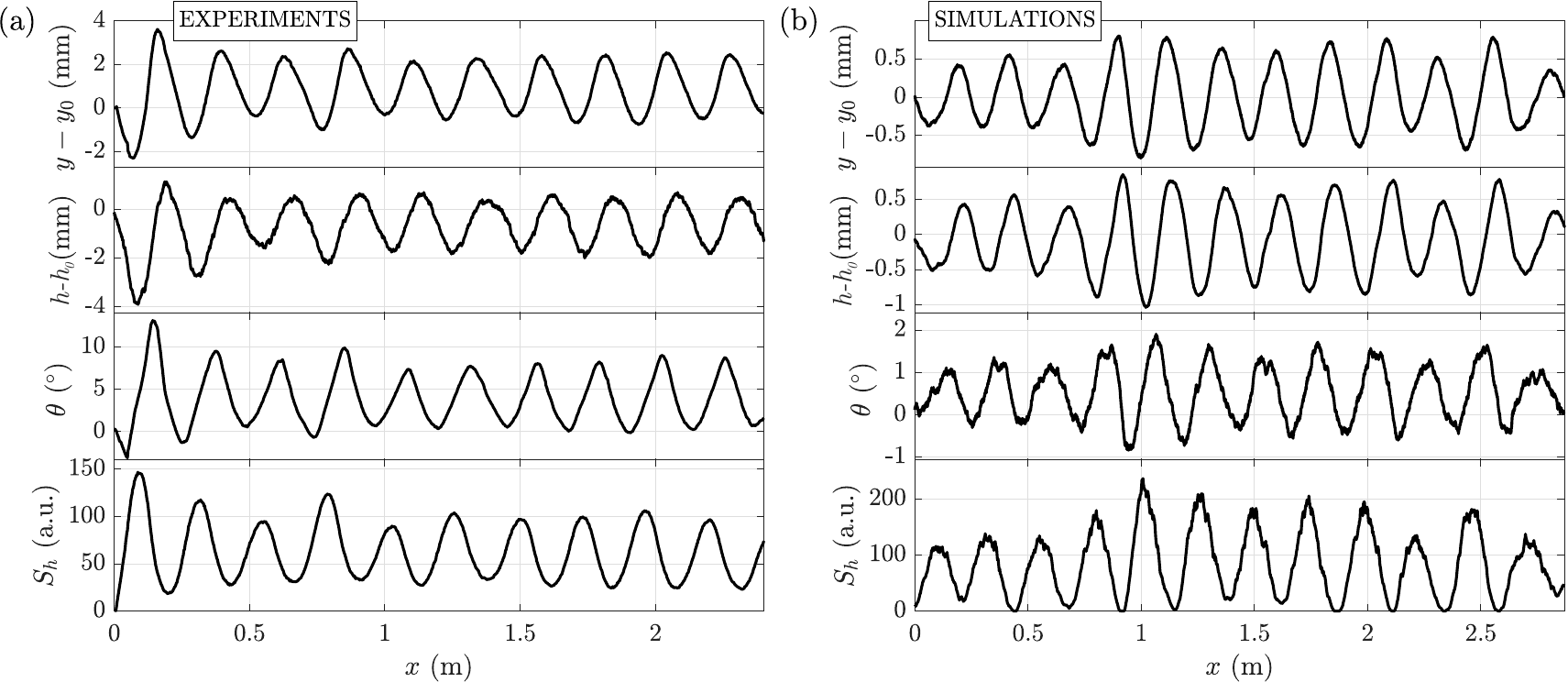}
\caption{\label{fig:ex_signals} Examples of surface periodic instabilities measured on the different signals. From top to bottom: slider altitude $(y - y_0)$, topography $(h-h_0)$ in its wake, slider pich angle $\theta$ and heap size $S_h$ (see text). The topography signal has been shifted to correspond to the rear corner of the slider.
(a) Experimental data [$m = 27$~g, $l = 30$~mm, $w = 40$~mm, $P = 220$~Pa, $\alpha=45$°]. 
(b) Numerical simulation [$m = 0.15$~g, $l = 48$~mm, $P = 61$~Pa, $\alpha=45$°].}
\end{figure*}

Figure~\ref{fig:sketch}(d) displays a snapshot of the simulation zoomed around the slider, the full simulation box length being of about $3$~m. The slider is made out of grains with diameter $1.2 d_p$.
The force on each grain is computed in the same way as any other grain, but instead of solving the equations of motion of every individual grain, the motion of the slider as a whole is integrated instead using the sum of all forces, the pulling force and its weight.
To apply the pulling force, a numerical spring is introduced between a virtual point moving forward at a constant speed and the point of application of the force. The stiffness of this spring does not have any influence on the results as long as it is taken high enough so that the global displacement is imposed at constant speed, as in experiments, and all fluctuations are negligible. All the numerical results presented here were obtained with $k=10$~N.m$^{-1}$.

In a typical simulation run, 100~000 grains are initially left to settle under their own weight on top of a monolayer of large grains fixed at the base corresponding to the bottom boundary condition. In order to avoid crystallization of a monodisperse medium, a uniform distribution of size (between $[0.8 d_p ; 1.2 d_p]$) was used. For each set of parameters, six initial states were generated (using six values of the seed in the random generator), leading to six macroscopically identical granular beds whose local arrangement differ. This allows one to check the sensitivity to the initial packing configuration and gives an estimate of the statistical spread of the results.

The values used in the simulation are chosen to typically match experimental values (see Table~\ref{tab:table1}). Yet, only a qualitative comparison should be expected since major differences still exist (among which a purely 2D system and ideally spherical grains in the simulation).

\section{\label{sec:results}Results}

\subsection{Observations}

Figure~\ref{fig:ex_signals} shows two typical examples (experimental and numerical) of the instability which occurs under a range of parameters which is in part investigated in Sec.~\ref{sec:results}B, \ref{sec:results}C and \ref{sec:results}D.
From top to bottom, the panels display the slider altitude $(y-y_0)$, the topography in its wake $(h-h_0)$, its pitch angle $\theta$ and the size of the heap in front of the slider $S_h$ (see Appendix~\ref{sec:appendix}).
All data sets show the instability from its initial state, where the slider simply rests horizontally on the granular layer, under its own weight. The topography ($h-h_0$), i.e., the pattern imprinted into the sand bed, is obviously measured after the slider has passed. The signal is thus shifted so that the $x$-position matches that of the rear angle of the slider and the topography starts at zero when $x=0$. 

In spite of the fundamental differences between our experimental setup and numerical simulations (3D vs. 2D, rough grains vs. ideal disks, grain stiffness etc) both data sets qualitatively display the same behavior. Indeed, in both cases, the instability is triggered as soon as the slider is dragged along the granular surface with no noticeable transient. The periodicity of all signals is very clear and fluctuations in their amplitudes are visible, with a strong correlation between the various physical quantities. Note that most signals show significant deviation from a harmonic function, possibly most evidently on the pitch angle $\theta$. This anharmonicity depends on the mechanical parameters of the system.

\begin{figure*}[t]
\includegraphics[width=0.98\textwidth]{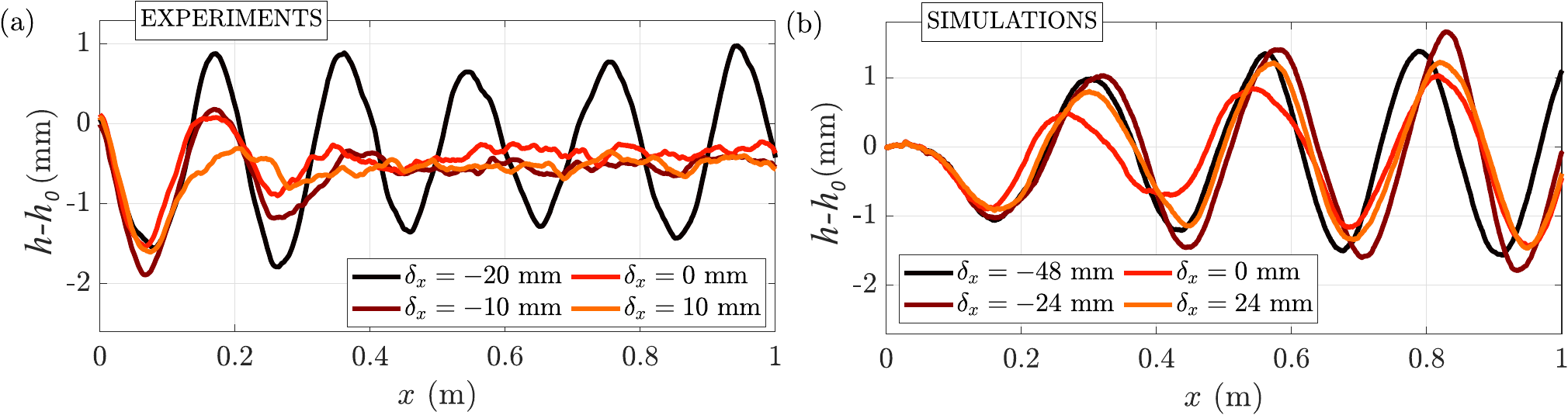}
\caption{\label{fig:deltax} 
Topography ($h-h_0$) as a function of the distance traveled $x$ displaying the role of the contact point horizontal coordinate $\delta_x$ in the appearance of the instability.
(a) Experiments [$m = 18.3$~g, $l = 30$~mm, $w=40$~mm, $P = 150$~Pa, $\alpha=45$°, $\delta_y = 10$~mm]. 
(b) Numerical simulations [$m = 0.08$~g, $l = 48$~mm, $P = 32$~Pa, $\alpha=45$°, $\delta_y = 12$~mm].
The effect of $\delta_x$ is stronger in experiments than in simulations.}
\end{figure*}

Let us denote $\langle . \rangle$ the average of the signal over the total distance traveled by the slider.
Experimentally, the average $\langle y-y_0 \rangle$ is always positive. This is due to the fact that $\langle \theta \rangle >0$ [Fig.~\ref{fig:ex_signals}(a)]. On the contrary, $\langle h \rangle <0$ [Fig.~\ref{fig:ex_signals}(a), second panel] as the slider pushed grains aside when going forward (3D experiment). This is not true in the 2D simulations, for which $\langle h \rangle \simeq 0$ [Fig.~\ref{fig:ex_signals}(b), second panel]. Note that in simulations, $\langle y-y_0 \rangle \simeq 0$ as the pitch angle $\theta$ always remains small [Fig.~\ref{fig:ex_signals}(b), first and third panels].
Moreover, one can see that the average pitch angle $\langle \theta \rangle$ is positive, meaning that the slider points upwards more than downwards. This remains true over the entire range of explored parameters, both in our simulations and our experiments. 
In the numerical example [Fig.~\ref{fig:ex_signals}(b), bottom panel], the heap formed in front of the slider has completely emptied at the end of each cycle ($S_h$ vanishes to zero), while on the other hand, the corresponding experimental data shows oscillations between two finite values. However, both behaviors can be observed either in our experiments or in our simulations depending notably of the mass or length of the slider.

The phase lag between the quantities at play is not trivial and may vary as the mechanical and geometrical properties are changed. However, the topography left behind the slider matches the trajectory of the rear corner, when the latter constantly remains in contact with the granular bed, which implies a geometrical relationship between the altitude, $y$, the pitch angle, $\theta$, the length, $l$ and the topography, $h$: 
\begin{equation}
h - h_0 = (y -y_0)– \frac{l}{2} \sin \theta  \,\, .
\end{equation}
Let us mention that when approaching the stable/unstable transition intermittency can occur. Oscillations can appear on and off irregularly during the course of an experimental or numerical run, and the system shows great sensitivity to small fluctuations in the initial state of the granular bed.

\subsection{Role of the contact point}
\label{sec:contact_point}

%\begin{figure*}[t]
%\includegraphics[width = .47\textwidth]{fig_3a}
%\includegraphics[width = .47\textwidth]{fig_3b}
%\caption{\label{fig:deltax} 
%Topography ($h-h_0$) as a function of the distance traveled $x$ displaying the role of the contact point horizontal coordinate $\delta_x$ in the appearance of the instability.
%(a) Experiments [$m = 18.3$~g, $l = 30$~mm, $w=40$~mm, $P = 150$~Pa, $\alpha=45$°, $\delta_y = 10$~mm]. 
%(b) Numerical simulations [$m = 0.08$~g, $l = 48$~mm, $P = 32$~Pa, $\alpha=45$°, $\delta_y = 12$~mm].
%The effect of $\delta_x$ is stronger in experiments than in simulations.}
%\end{figure*}

Let $\delta_x$ and $\delta_y$ be the relative coordinates of the contact point where the force is applied to the slider, the origin being located at the center of the bottom plate of the slider. For instance on Fig.~\ref{fig:sketch}(c), $\delta_x = -20$~mm and $\delta_y = 10$~mm. Figure~\ref{fig:deltax} shows the role of $\delta_x$ both in the experiments and simulations (for a given value of $\delta_y$). Experimentally, out of the four examples shown, only the experiment with $\delta_x = -20$~mm  leads to sustained oscillations in the topography [Fig.~\ref{fig:deltax}(a)]. In the three other examples, the oscillations are damped and quickly die out. In other words, experimentally the instability is only triggered when the slider is pushed from a point which is near it rear end. Numerically, the role of $\delta_x$ appears less important, and we report sustained oscillations in all cases, with little difference in amplitude in the range considered [Fig.~\ref{fig:deltax}(b)].
The origin of the major discrepancy between numerical and experimental results remains unclear and is briefly discussed in Sec.~\ref{sec:discussion}. 

\begin{figure}
\includegraphics[width = 0.49\textwidth]{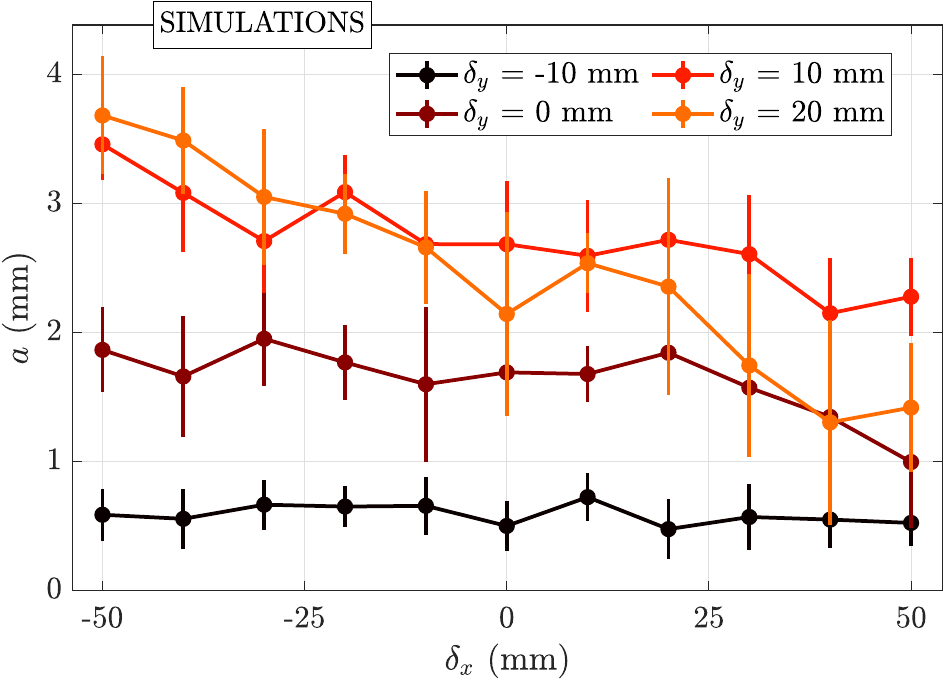}
\caption{\label{fig:deltaxy} 
Amplitude of the surface instability $a$ as a function of the horizontal location of the contact point $\delta_x$, for different vertical position of this contact point $\delta_y$ [$m = 0.08$~g, $l = 48$~mm, $P = 33$~Pa, $\alpha=45$°].}
\end{figure}

Figure~\ref{fig:deltaxy} shows the evolution of the amplitude $a$ of the topography as a function of $\delta_x$ for various vertical position $\delta_y$ of the contact point. Only numerical results are presented here, as the simulations make possible to widely extend the range of $\delta_x$ and $\delta_y$. Indeed, the force can virtually be applied anywhere in simulations, including on a point located inside the granular bed ($\delta_y < 0$), which is impossible in experiments. Negative values of $\delta_y$ (for instance $\delta_y=-10$~mm, Fig.~\ref{fig:deltaxy}) always lead to a stable motion (the resulting amplitude corresponding to fluctuations). In that case, whichever $\delta_x$, the positive torque exerted by the force tilts the slider counterclockwise, i.e., upwards. As a result, the front corner of the slider loses contact with the granular bed, which prevents the formation of a heap in front of the slider. Only partially resting on its bottom plate, the slider glides above the grains with only small fluctuations in $\theta$. A similar behavior also happens for high positive values of $\delta_y$ (not shown on Fig.~\ref{fig:deltaxy}), for which the slider is tilted clockwise, sliding on its front spatula. The instability is therefore observed for intermediate, null or positive values of $\delta_y$ only, for which  the topography resembles the periodic signals shown in Fig.~\ref{fig:deltax}(b).

One may expect that pushing the slider from behind ($\delta_x < 0$) should be unstable while pulling it from the front ($\delta_x > 0$) should stabilize it. Indeed, when pushed from behind, if the slider points upwards (which a positive pitch angle), then the applied external force creates a positive torque, which should increase the pitch further. Similarly, a negative pitch angle should be amplified. On the other hand, when pulled from the front, an upward-pointing slider should experience a negative torque and should pitch down. 
This is only somewhat true experimentally, $\delta_x = -20$~mm is indeed unstable but $\delta_x = -10$~mm is surprisingly stable. 
Numerically, for all positive values of $\delta_y$, the amplitude decreases slightly with increasing $\delta_x$. But even more importantly, even in the case where $\delta_y =0$, the system is always unstable for all values of $\delta_x$. 

The torque applied by the pushing force can easily be computed and varies over time as both the pulling force and the pitch angle vary. However, averaged over an entire wavelength, the torque is positive for $\delta_y >0$. 
Moreover, for a given value of $\delta_y$ the average torque is expected to be identical for a value of $\delta_x$ and $-\delta_x$ (since in the simulations the center of mass of the slider is located in the middle of the bottom plate).  Yet, as can be seen in Fig.~\ref{fig:deltaxy} (see $\delta_y = 20$~mm and $\delta_x = \pm 50$~mm for a clear example), equal values of the averaged applied torque can lead to very different amplitudes in the oscillations. 
A way to bias the torque is to choose a nonuniform mass distribution of the slider. We have tested this idea in our experimental setup and found that indeed, an asymmetrical mass distribution has a strong effect on the instability. Shifting the weight forward (closer to the front spatula) has a clear tendency to hinder the oscillations, while shifting it back only seems to favor large oscillations.

\subsection{Effect of the slider mass and length}

\begin{figure*}[t]
\includegraphics[width = \textwidth]{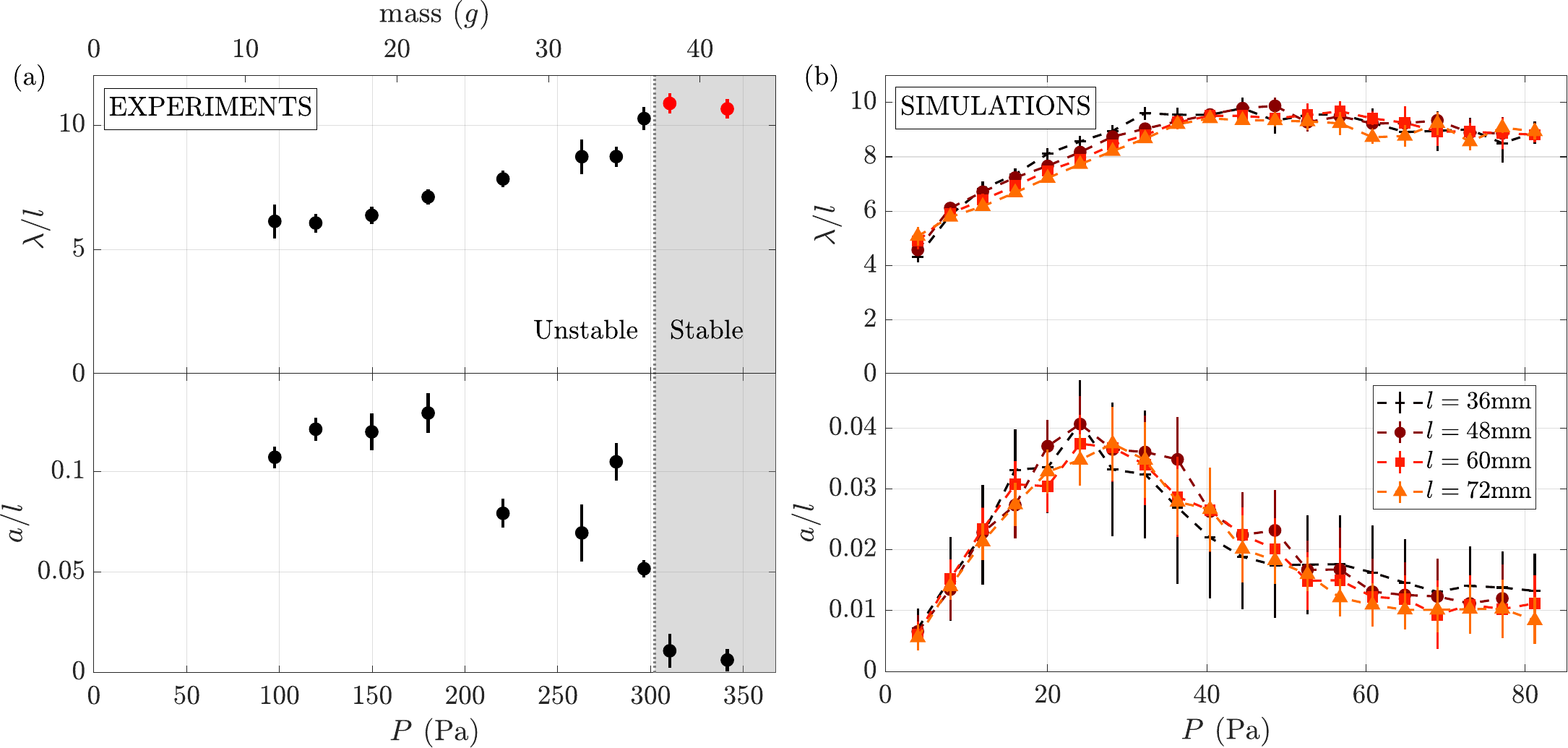}
\caption{\label{fig:ml}
Normalized wavelength $\lambda/l$ (top panels) and amplitude $a/l$ (bottom panel) of the surface instability as a function of the confining pressure. (a) Experiments. The top axis indicates the slider mass $m$. The red points for the wavelength in the stable regime (gray region) are obtained from damped oscillations (see text). Vertical bars indicate the error bars (when not visible, they are smaller than the dot size) [$l=30$~mm $w=40$~mm, $\alpha=45$°, $\delta_x = -20$~mm, $\delta_y = 10$~mm].
(b) Simulations [$\alpha=45$°, $\delta_x = 0$~mm, $\delta_y = 0$~mm].}

\end{figure*}

Figure~\ref{fig:ml} presents the evolution of the wavelength and amplitude normalized by the slider length, $\lambda/l$ (top panels) and $a/l$ (bottom panels), as a function of the confining pressure $P=mg/lw$ for both the experiments and numerical simulations. Experimentally, increasing the confining pressure $P$ leads to an increase in wavelength and a decrease in amplitude. When the confining pressure $P$ is high enough, the periodic instability disappears and the slider motion becomes stable [Fig.~\ref{fig:ml}(a), gray region]. In that case, the amplitude drops to zero. Interestingly, even in the stable regime, one can extract a wavelength by forcing the slider out of equilibrium by briefly applying a vertical downward force on its front spatula. One can then observe a few damped oscillations before the slider reaches a stable motion. The wavelengths mesured in this damped regime is shown as red dots in the stable regions [Fig.~\ref{fig:ml}(a), top panel]. They nicely follow the tendency found in the unstable zone, and display a possible plateau, although more data would be needed to conclude confidently.

In the experiments, the lightest slider weighs $m=11$~g (size $l=30$~mm, $w=40$~mm). Lowering the confining pressure even further is difficult, as the slider must have a proper mechanical rigidity, a layer of grains glued under it, and a metal cylinder on top for contact with the metal plate. For this slider, we were not able to probe confining pressures $P$ below 90~Pa [Fig.~\ref{fig:ml}(a)]. Numerical simulations however allow one to explore an extended range of confining pressure, in particular towards the lower values. Figure~\ref{fig:ml}(b) displays the normalized wavelength $\lambda/l$ and amplitude $a/l$ of the instability as a function of the confining pressure $P$, for different slider length $l$. All data collapse on a master curve, indicating that the resulting pattern scales as the slider length $l$. The smallest values of the confining pressure lead to oscillations with $\lambda \simeq 5 l$ and a very small amplitude (about $1\%$ of $l$). 
The wavelength increases with increasing pressure and seems to reach a plateau. The amplitude reaches a maximum (around $P=25$~Pa) before decreasing towards a plateau at higher pressure. Although the experimental pressure range is significantly different from the numerical one, experiments and simulations are in good qualitative agreement, both displaying (1) the instability amplitude decrease for an increase confining pressure (heavier or shorter slider) and (2) an increase in the wavelength until reaching a plateau.
In both cases, there seems to be an optimal value of the confining pressure for the instability. We can propose the following interpretation. On the one hand, when the confining pressure is too small, the slider is too light to scrape the grains and slips over the surface without forming a heap. On the other hand, for large pressure the slider accumulates a large heap but seems too heavy to rise above the pile and remains in a sunken state. In both cases the instability vanishes. The existence of an optimal value for the confining pressure is reminiscent of antlion traps which exhibit an optimum of the probability of an ant's capture as a function of its mass \cite{Humeau2015}, interpreted as the existence of a minimum in the friction coefficient as a function of the applied pressure \cite{Crassous2017}.

\subsection{Effect of the spatula front angle}
\label{sec:angle}

\begin{figure}
\includegraphics[width = \columnwidth]{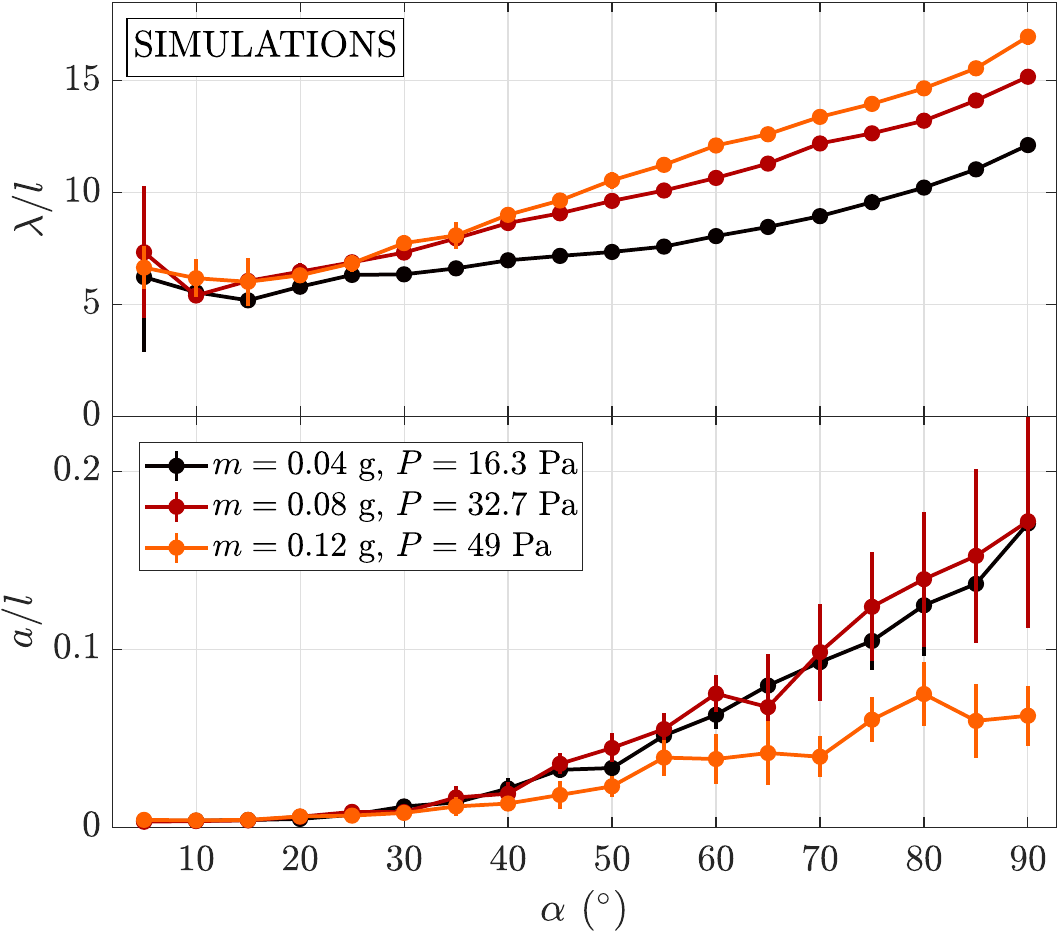}
\caption{\label{fig:alpha} Normalized wavelength $\lambda/l$ and amplitude $a/l$ as a function of the slider spatula front angle $\alpha$ in simulations. [$l=48$~mm, $\delta_x = 0$~mm, $\delta_y = 0$~mm]}
\end{figure}

Another important parameter is the slider spatula front angle, denoted $\alpha$ (see Sec.~\ref{sec:exp_methods}). Indeed, varying $\alpha$ directly affects the scraping of the granular bed, as well as the direction of the forces exerted by the grains on the spatula. Increasing the spatula front angle, $\alpha$, leads to a more efficient scraping ({\it bulldozer-like effect}) and the formation of a larger heap. Consequently, the pattern wavelength and amplitude both increase (Fig.~\ref{fig:alpha}). This remains true for a wide range confining pressures: three values corresponding approximately to the growth, peak and decrease in amplitude for $\alpha = 45$° in Fig.~\ref{fig:ml}(b) are shown. 

Interestingly, both the amplitude and wavelength keep increasing when $\alpha$ is increased for $\alpha>60$°. However, when performing numerical scraping tests at fixed $\theta = 0$°, $y = y_0$ where $y_0$ is taken slightly lower than the surface level of the granular bed, we noticed that the rate of collection of grains in the heap, or scraping efficiency (the proportion of the scraped surface that ends up in the heap instead of going under the slider), increases from $\alpha = 10$° to $\alpha = 60$° where it reaches a plateau -- the growth rate did not differ significantly between $60$° and $90$°. In the free simulation, scraping is not only affected by the depth of the slider and $\alpha$ but also by its vertical speed, pitch and rotation rate. However, the heap growth rate also seems to saturate for $\alpha \geq 60$° while the wavelength and amplitude keep increasing. Understanding the dynamic scraping requires further investigation and might explain why the amplitude and wavelength keep increasing for $\alpha \geq 60$°.

For high values of the amplitude, the effect of the finite thickness of the granular bed cannot be neglected. Indeed, the thickness of the granular bed is $8.4$~mm and the amplitude of the pattern reaches half this value when $a/l=0.087$, indicating that the data points for $\alpha \geq 60$° for $P\geq 30$~Pa on Fig.~\ref{fig:alpha} may be affected by boundary effects. One has therefore to keep in mind that an interaction with the bottom  boundary condition in the numerical simulations, corresponding to a monolayer of fixed grains, can exist, at least for high values of $\alpha$. Comparison with simulations using a thicker granular bed could help quantifying this effect.

\section{Discussion}
\label{sec:discussion}

Similarly to numerical simulations, the stiffness of the plate used to pull the slider in the experiment does not play any role neither in the formation of the pattern, nor on its amplitude and wavelength, as long as it is high enough to keep its deflection small (typically of the order of a grain size) and ensure a horizontal traction force. Varying the metal plate stiffness down to 500~N.m$^{-1}$ did not change the results neither qualitatively nor quantitatively. However, having both the translational and rotational degrees of freedom is critical to obtain this instability. Blocking one degree of freedom leads to the disappearance of the instability.

As discussed in Sec.~\ref{sec:contact_point}, the position of the contact point has a fundamentally different effect in simulations and experiments. As explained, the torque exerted by the traction force -- computed relative to the center of gravity of the slider (which is located at $\delta_x = 0$) -- has a stabilizing effect on $\theta$ when $\delta_x > 0$, and a destabilizing effect when $\delta_x < 0$, analogous to the stability (respectively instability) of a trailer when the car is going forward (respectively backward).
While this stabilizing or destabilizing effect can partly explain the decrease in amplitude when $\delta_x$ is increased [see Fig.~\ref{fig:deltax}(b)], it is not enough to understand the instability since sustained oscillations are observed both in the push and pull positions (in simulations).
As a reminder, on the other hand, in our experimental setup the instability entirely disappear if the slider is pulled  (i.e., $\delta_x>0$) rather that pushed. 

The origin of this major discrepancy between experiments and simulations remains to be identified. There are several inherent differences between our numerical simulations and experimental setup, for instance 2D vs 3D, ideal vs real grains, or a large difference in particle rigidity.
Moreover, numerically a purely horizontal force is applied to push the slider. 
Experimentally, a stiff plate pushes onto a horizontal cylinder, but as the slider moves up or down, friction at the contact point induces a vertical component to the pushing or pulling force. We have tried to minimize friction at the contact point by polishing and lubricating the cylinder and metal plate surfaces, or instead to increase the friction by gluing a piece of fine sandpaper on one or both surfaces. Clearly, this can affect the onset of the instability but no general trends were found. Increasing the friction can sometimes kill an otherwise unstable pattern but oscillations can still be found using the sand paper contact for other parameters. 

We can also reduce the mobility of the contact point by adding a small amount of water, whose high surface tension causes the surfaces to adhere to one another through a capillary bridge. Translating the cylinder vertically along the metal plate then requires exceeding a force threshold needed to make the contact lines recede and/or progress. Again, this is sufficient to kill the instability in some cases, while oscillations can still be observed using other parameters. 

The mechanism by which the instability can be hindered is not clear yet, but it appears that introducing a vertical component to the traction force while impeding the motion of this contact point can have a strong effect on the stability of the system. In an attempt to reduce the effect of friction, we tried pulling the slider using a string. One end of the string is glued to the cylinder, and the other far from the slider (forward). The string is kept under tension and remains horizontal, which imposes a purely horizontal pulling force. Surprisingly, using this pulling method, we were unable to observe sustained oscillations. For a set of parameters which leads to a well-defined instability, replacing the cylinder-plate contact by a string instantly kills the instability. 

\section{Conclusion}

We report a surface instability leading to pattern formation reminiscent of similar processes in hydrodynamics or grains \cite{Hewitt2012a}. However, unlike the washboard instability or other inertia-driven phenomena such as the speed wobble \cite{Sharp2004,Plochl2012,Rosatello2015}, snaking of a car-trailer \cite{Korayem2022, Darling2009} or porpoising of boats \cite{Thornhill2000,Masumi2017,Ikeda2000,Sun2011}, the instability reported is not sensitive to the value of $v$ in the range considered (Table~\ref{tab:table1}) as it occurs in the quasistatic regime. For this reason, it seems more promising to attempt to model this instability based on geometrical considerations rather than a force-based model relying on time derivatives. However, previous works \cite{Guillard2014, Gravish2010} have used force measurements to describe the quasistatic motion of an intruder (cylinder or plow) in a granular bed. Their results could provide inspiration to model our instability. This deserves further investigation and will be the focus of a future paper.
Indeed, our DEM simulations easily allow one to quantify the forces and velocities distributions, which could provide additional insights in the origin of the instability (see Appendix~\ref{sec:FV}).

Finally, it appears that experimentally, the details of the mechanics of the contact point between the slider and the object that pushes it are crucial. The exact nature of the contact forces which annihilates the instability remains to be understood and deserves further work.

\begin{acknowledgments}

The authors acknowledge two anonymous referees for their pertinent comments which greatly improved the paper.

\end{acknowledgments}

\appendix

\section{Image analysis and data processing}
\label{sec:appendix}

The altitude $y$ of the slider center of mass and its pitch angle $\theta$ are computed from both the experimental records and numerical simulations. Images from the experiments are analyzed by a homemade Matlab software which detects the pattern (dots) printed on the slide of the slider (Sec.~\ref{sec:exp_methods}). 
Based on this detection, the program computes the slider altitude, $y$, and its pitch angle, $\theta$. We also quantify the heap of grains carried in front of the slider. In numerical simulations, the heap size is estimated by counting the number of grains moving with a velocity above an arbitrary threshold of 25~mm/s, corresponding to $62.5\%$ of the traction speed ($v=40$~mm/s, see Table~\ref{tab:table1}). Any other choice of that threshold between $50\%$ and $75\%$ yields an equally satisfactory detection and the velocity threshold value was chosen to best fit the expected geometry of the heap. In experiments, the heap is quantified as the surface encompassed between the slider front, the prolongation of the slider bottom and the heap front edge, this latter being found by simple edge detection (Fig.~\ref{fig:sketch}(c)). In spite of the difference in quantification and unit (number of grains in simulations vs. surface in mm$^2$ in experiments), both quantities are representative of the variation of the same variable which is named afterwards $S_h$, as the heap size, and indicated in arbitrary units. 

\section{Forces and Velocities}
\label{sec:FV}

\subsection{Force distribution}

\begin{figure}[t]
\includegraphics[width = \columnwidth]{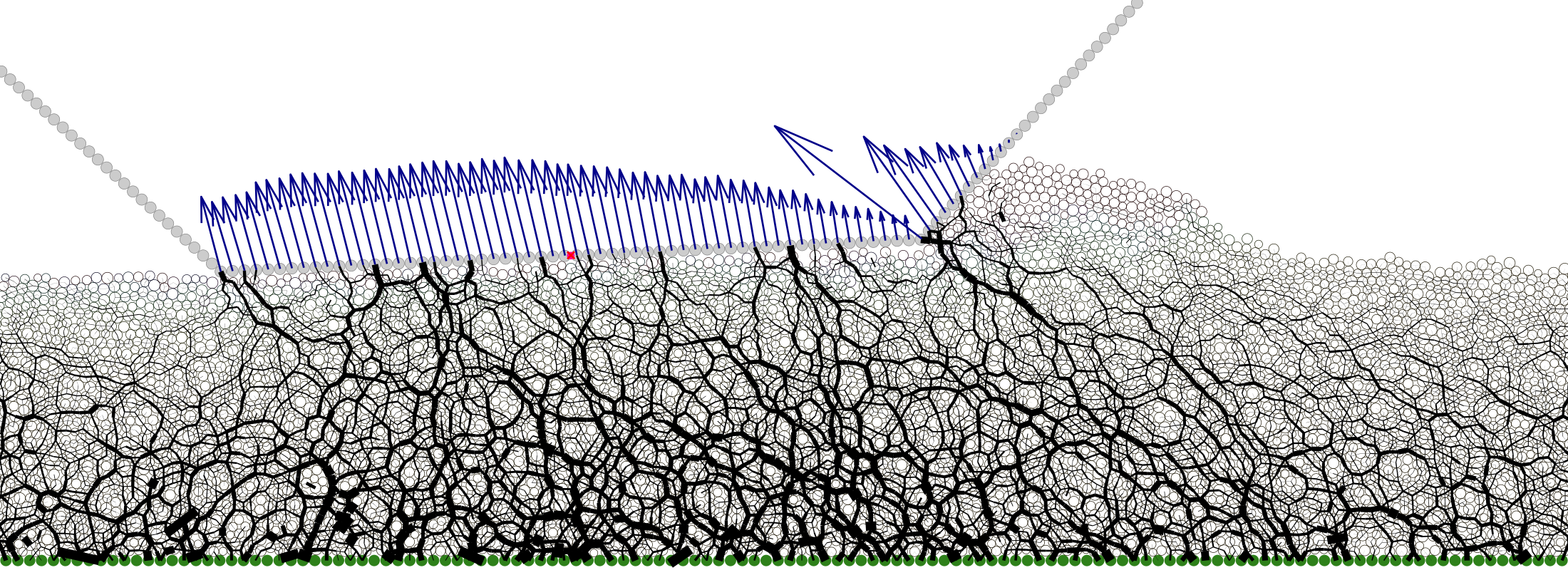}
\caption{\label{fig:forces} Snapshot of a numerical simulation showing the forces between grains and on the slider. Instantaneous contact forces are shown in black while an averaged force profile is represented by the blue arrows on the slider bottom and front spatula  [$l=36$~mm, $\delta_x = 0$~mm, $\delta_y = 0$~mm].}
\end{figure}

Since numerical simulations give access to all the contact forces, one can easily investigate their distribution both within the granular bed, and between the grains and the slider, as represented on Fig.~\ref{fig:forces}. These results might help shed light on the origin of the instability.

For each pair of grains in contact, a black line whose thickness scales as the contact force is drawn between their centers, thus showing the global instantaneous configuration of force chains (Fig.~\ref{fig:forces}). One can notice that far away from the slider a typical hydrostatic-like configuration with no preferred angle is observed. In the vicinity of the slider, and particularly near its front spatula, strong force chains span the entire gap between the slider surface and the bottom of the granular layer, and show a preferential direction. A proper statistical analysis of the direction of these forces depending on the conditions is outside the scope of this paper but would deserve further attention.

Figure~\ref{fig:forces} also displays the local average force (blue arrows) experienced  by the slider (made of individual grains held together as a solid body). On the front spatula, the force decreases roughly linearly from the front corner towards the free surface. On the contrary, on the bottom plate, the force increases when scanning from the front to the rear corner. While it increases rapidly, the force tends to saturate to reach a plateau over a large part of the bottom plate. 

The front corner grain, which belongs both to the bottom and front plates, seems to play a singular role. Yet, it is worth noting that as the pitch angle of the slider are not fixed in this simulation, the forces acting on the slider may vary in time, which might explain the averaged high value for the front corner grain. 

\subsection{Velocity distribution}

\begin{figure}[t]
\includegraphics[width = \columnwidth]{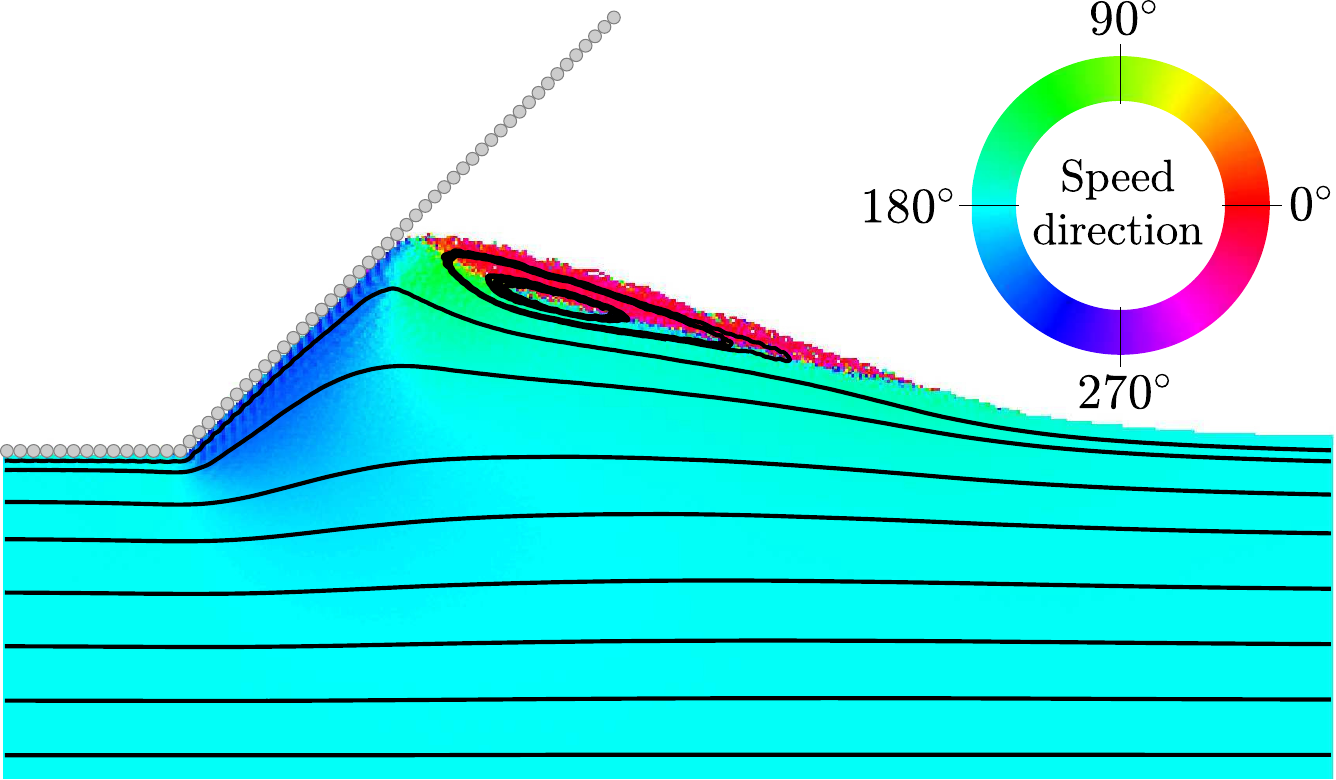}
\caption{\label{fig:direction} Granular flow in the moving frame of reference on the slider, in a rotation-locked simulation. The black lines are streamlines, showing a vortex close to the free surface. The colormap indicates the local direction of the flow.}
\end{figure}

Using our simulation, we have also investigated the flow within the granular heap formed in front of the slider, as illustrated by Fig.~\ref{fig:direction}. In order to obtain smooth data, we chose to simulate a situation in which the tilt angle was set to zero. Yet, the overall behavior remains similar when the angle is free to vary. 
Figure~\ref{fig:direction} shows the streamlines (black lines), and the colormap indicates the direction of the flow, in the frame of reference of the slider.
A clear forward motion (in red) is visible near the free surface of the dragged pile, and a recirculating vortex can be observed within the dragged pile.
The figure reveals the structure of the granular flow which clearly deserves further investigation.

%\bibliography{biblio}

%apsrev4-2.bst 2019-01-14 (MD) hand-edited version of apsrev4-1.bst
%Control: key (0)
%Control: author (8) initials jnrlst
%Control: editor formatted (1) identically to author
%Control: production of article title (0) allowed
%Control: page (0) single
%Control: year (1) truncated
%Control: production of eprint (0) enabled
%

\end{document}